\documentclass[sigconf]{acmart}
\AtBeginDocument{%
  }

\setcopyright{none}
\acmDOI{10.48550/arXiv.2508.07454}
\acmConference[LIMITS '25]{11th Workshop on Computing Within Limits}{June 26--27, 2025}{Online}
\acmISBN{}

\usepackage{array}

\begin{document}

%%
%% The "title" command has an optional parameter,
%% allowing the author to define a "short title" to be used in page headers.
\title{An Empirical Inquiry into Surveillance Capitalism: Web Tracking}

%%
%% The "author" command and its associated commands are used to define
%% the authors and their affiliations.
%% Of note is the shared affiliation of the first two authors, and the
%% "authornote" and "authornotemark" commands
%% used to denote shared contribution to the research.
\author{Nils Bonfils}
%\authornote{Both authors contributed equally to this research.}
\email{nils.bonfils@mail.utoronto.ca}
\orcid{0009-0000-1673-8606}
%\orcid{}
\affiliation{%
  \institution{University of Toronto}
  \city{Toronto}
  \state{Ontario}
  \country{Canada}
}

%%
%% By default, the full list of authors will be used in the page
%% headers. Often, this list is too long, and will overlap
%% other information printed in the page headers. This command allows
%% the author to define a more concise list
%% of authors' names for this purpose.
\renewcommand{\shortauthors}{Bonfils}

%%
%% The abstract is a short summary of the work to be presented in the
%% article.
\begin{abstract}
  The modern web is increasingly characterized by the pervasiveness of Surveillance Capitalism. This investigation employs an empirical approach to examine this phenomenon through the web tracking practices of major tech companies --- specifically Google, Apple, Facebook, Amazon, and Microsoft (GAFAM) --- and their relation to financial performance indicators. Using longitudinal data from WhoTracks.Me spanning from 2017 to 2025 and publicly accessible SEC filings, this paper analyzes patterns and trends in web tracking data to establish empirical evidence of Surveillance Capitalism's extraction mechanisms. Our findings reveal Google's omnipresent position on the web, a three-tier stratification among GAFAM companies in the surveillance space, and evidence suggesting an evolution of tracking techniques to evade detection. The investigation further discusses the social and environmental costs of web tracking and how alternative technologies, such as the Gemini protocol, offer pathways to challenge the extractive logic of this new economic order. By closely examining surveillance activities, this research contributes to an ongoing effort to better understand the current state and future trajectory of Surveillance Capitalism.
\end{abstract}

%%
%% The code below is generated by the tool at http://dl.acm.org/ccs.cfm.
%% Please copy and paste the code instead of the example below.
%%
% \begin{CCSXML}
% <ccs2012>
%  <concept>
%   <concept_id>00000000.0000000.0000000</concept_id>
%   <concept_desc>Do Not Use This Code, Generate the Correct Terms for Your Paper</concept_desc>
%   <concept_significance>500</concept_significance>
%  </concept>
%  <concept>
%   <concept_id>00000000.00000000.00000000</concept_id>
%   <concept_desc>Do Not Use This Code, Generate the Correct Terms for Your Paper</concept_desc>
%   <concept_significance>300</concept_significance>
%  </concept>
%  <concept>
%   <concept_id>00000000.00000000.00000000</concept_id>
%   <concept_desc>Do Not Use This Code, Generate the Correct Terms for Your Paper</concept_desc>
%   <concept_significance>100</concept_significance>
%  </concept>
%  <concept>
%   <concept_id>00000000.00000000.00000000</concept_id>
%   <concept_desc>Do Not Use This Code, Generate the Correct Terms for Your Paper</concept_desc>
%   <concept_significance>100</concept_significance>
%  </concept>
% </ccs2012>
% \end{CCSXML}

% \ccsdesc[500]{Do Not Use This Code~Generate the Correct Terms for Your Paper}
% \ccsdesc[300]{Do Not Use This Code~Generate the Correct Terms for Your Paper}
% \ccsdesc{Do Not Use This Code~Generate the Correct Terms for Your Paper}
% \ccsdesc[100]{Do Not Use This Code~Generate the Correct Terms for Your Paper}

%%
%% Keywords. The author(s) should pick words that accurately describe
%% the work being presented. Separate the keywords with commas.
%% A "teaser" image appears between the author and affiliation
%% information and the body of the document, and typically spans the
%% page.
% \begin{teaserfigure}
%   \includegraphics[width=\textwidth]{sampleteaser}
%   \caption{Seattle Mariners at Spring Training, 2010.}
%   \Description{Enjoying the baseball game from the third-base
%   seats. Ichiro Suzuki preparing to bat.}
%   \label{fig:teaser}
% \end{teaserfigure}

% \received{20 February 2007}
% \received[revised]{12 March 2009}
% \received[accepted]{5 June 2009}

%%
%% This command processes the author and affiliation and title
%% information and builds the first part of the formatted document.
\maketitle

% update figure 4 and desc, along with findings and paragraph in conclusion
% Add a sentence in limitations about why no ad rev for aapl and msft
% Env cost in discussion
% (expand on business model of Apple and Microsoft)

\section{Introduction}
Contemporary society is increasingly characterized by the pervasiveness of Surveillance Capitalism, a new economic order where human experience is claimed as free raw material for hidden commercial practices of extraction, prediction, and sales. While fundamentally reshaping digital interactions, this phenomenon remains conceptually elusive and difficult to grasp in concrete terms. This elusiveness calls for empirical investigation to understand its actual mechanisms and impact.

Surveillance Capitalism originated at Google, when the corporation discovered that users' seemingly inconsequential digital traces, also called “data exhaust”, could be algorithmically processed and analyzed to predict future behavior and monetized through targeted advertising \cite{zuboff_big_2015}. The internet, particularly the World Wide Web, provided the essential infrastructure for this new economic logic to flourish. The web's client-server architecture, coupled with lagging regulatory frameworks, created optimal conditions for surveillance operations to take root without restraints.

The growing trends towards the use of web applications over more traditional “native” applications has accelerated this transformation. With individuals increasingly migrating daily activities to web applications, the web has evolved from a document-sharing medium to an application platform. This consolidation of activity on the web provides an ideal environment for surveillance operations, enabling tech companies to monitor user behavior across various services through web tracking.  This surveillance power is further concentrated as these same companies provide access to the web platform through their browsers, with Google, Apple, and Microsoft’s web browsers representing approximately 90\% of all browser usage.

Web tracking refers to the collection of data about users' online activities, including pages visited, clicks made, time spent on sites, and numerous other behavioral signals. These collection mechanisms range from simple cookies to sophisticated fingerprinting techniques capable of identifying users across multiple devices and sessions. While some tracking serves legitimate purposes like authentication and site functionality, the vast majority enables the surveillance apparatus that powers contemporary Surveillance Capitalism. Both direct outcomes (targeted advertising) and indirect outcomes (product improvement) of this surveillance enable the creation of “behavioral prediction products” that have tangible monetary value.

Web tracking data offers researchers a valuable opportunity to investigate the materialization of Surveillance Capitalism. By analyzing patterns and evolution in tracking technologies, we can begin to map the actual extent and mechanisms of surveillance on the web, moving beyond theoretical frameworks to empirical evidence. This methodological approach allows for critical assessment of surveillance practices across digital ecosystems and their relationship to corporate financial performance.

This investigation will seek to establish empirical evidence of Surveillance Capitalism by analyzing web tracking practices of major tech companies --- specifically Google, Apple, Facebook, Amazon, and Microsoft (GAFAM) --- and their relation to financial performance indicators. Three primary research questions will be addressed:

\begin{itemize}
\item {\textbf{RQ1}}: Within GAFAM, which entities demonstrate the highest engagement in surveillance practices?
\item {\textbf{RQ2}}: Do web tracking metrics correlate with the entity's advertising revenue?
\item {\textbf{RQ3}}: Can web tracking provide quantitative evidence of Surveillance Capitalism activity?
\end{itemize}

\section{Background}
This work builds on the theoretical framework of Surveillance Capitalism established by Zuboff \cite{zuboff_big_2015, zuboff_age_2019}. Most contributions based on this framework are predominantly qualitative in nature \cite{landwehr_problems_2023, power_theorizing_2022}. Few empirical studies directly reference Surveillance Capitalism \cite{amiel_mapping_2023}.

The dearth in empirical studies can be attributed to the active concealment of activities and practices by corporations partaking in Surveillance Capitalism. Entities that surveil also actively conceal their data collection practices to maintain this asymmetrical access to information that is essential for Surveillance Capitalism to function. In Zuboff's seminal book, she defines Surveillance Capitalism as an economic order, an economic logic, a mutation of capitalism, a foundational framework, a threat, the origin of a new power, a movement and an expropriation of human rights. \cite{zuboff_age_2019} Despite this rich definition, such a concept remains difficult to operationalize, making it challenging to identify empirically. Additional work is needed on defining Surveillance Capitalism, but more importantly on identifying ways to evaluate and measure its different aspects. However, empirical researchers investigating this topic implicitly place themselves in opposition to the dominant entities of this new economic order. 

Nonetheless, there has been a wealth of empirical research examining web trackers, ad blockers, and privacy enhancing technologies outside of the Surveillance Capitalism theoretical framework. Previous research on web tracking has focused on specific countries \cite{bailey_look_2019}, continents \cite{helles_infrastructures_2020}, or regulatory landscapes \cite{johnson_privacy_2023}. Some studies adopt a more global perspective \cite{samarasinghe_towards_2019}. However, most of those studies only provide static snapshots of the web tracking landscape. Few studies conduct longitudinal analyses \cite{lerner_internet_2016, lukic_impact_2024} and fewer systematically link trackers to their parent corporate entities. This investigation uses data from a long-standing open source database that allows investigation of web-tracking over multiple years \cite{karaj_whotracks_2019}.

This study presents a significant opportunity for interdisciplinary research that bridges empirical research of privacy-enhancing technologies with the theoretical framework of Surveillance Capitalism.

\section{Approach}
This investigation employs an empirical approach to examine the role of GAFAM companies in the web tracking space. Two primary data sources were used:

\begin{enumerate}
    \item WhoTracks.Me: an open source database of web trackers.
    \item GAFAM’s publicly accessible U.S. Securities and Exchange Commission (SEC) filings.
\end{enumerate}

Both datasets were integrated to examine the interplay between tracking practices and financial performance. 

\subsection{WhoTracks.Me Dataset}
WhoTracks.Me is a website and database that document web tracking activities \cite{karaj_whotracks_2019}. It originated within the Cliqz company in 2017 after the acquisition of the Ghostery web browser extension\footnote{\url{https://www.ghostery.com/ghostery-ad-blocker}}. While Cliqz, a company focused on creating both a privacy-oriented web browser and search engine, discontinued operations, the work on Ghostery extension and associated database continues to receive monthly updates.

 Data collection occurs through “real” web browsing by Ghostery users who have consented to share their tracking data during their browsing. This dataset is self-described as “the largest and longest measurement of online tracking” with data dating back to 2017. This gives a unique opportunity to study the evolution of web tracking over time.\footnote{\url{https://github.com/whotracksme/whotracks.me}}

WhoTracks.Me data is a publicly accessible AWS S3 bucket. As of April 2025, there are approximately 29 gigabytes of tracker data available. The data is structured hierarchically. At the top, directories represent each month of data collected. The monthly directory is further divided by region (e.g. “Global”, “US”, “EU”). Each region corresponds to the place the data was collected from which enables comparative analysis across regulatory environments. Given the transnational nature of Surveillance Capitalism, this research will focus on the “Global” region. 

Within the regions’ folders, there are five comma-separated value (CSV) files. Each file aggregates data around a specific entity in the web tracking ecosystem. This investigation will focus on the “companies.csv” dataset.\footnote{For an explanation of the other datasets available, the reader is encouraged to consult the database’s documentation directly: \url{https://github.com/whotracksme/whotracks.me/blob/master/whotracksme/data/Readme.md\#datasets} } This dataset aggregates web tracking data about the top companies and Table \ref{tab:vars} describes the relevant variables within that dataset.

\begin{table*}[h]
  \caption{Description of the relevant variables in the "companies.csv" dataset}
  \label{tab:vars}
  \begin{tabular}{|m{0.15\textwidth} m{0.6\textwidth} m{0.12\textwidth}|}
    \toprule
    Variable&Description&Possible values\\
    \midrule
    \hline
    reach & Proportional presence across all page loads (i.e. if a tracker is present on 50 out of 1000 page loads, the reach would be 0.05). & Floating point between 0 and 1\\
    \hline
    site\_reach & Presence across unique first party sites. e.g. if a tracker is present on 10 sites, and there are 100 different sites in the database, the site reach will be 0.1. \textbf{Important note}: In February 2019, this measure was redefined to the number of sites in the top 10,000 which have this tracker on more than 1\% of page loads. To stay consistent with the previous definition of that measurement, that value is divided by 10,000. & Floating point between 0 and 1\\
    \hline
    trackers & Average number of trackers present on the sites that uses at least one of this company’s trackers. & Positive floating point\\
    \hline
    content\_length & Average \textit{ Content-Length} HTTP headers received by third-party requests to trackers’ domains owned by this company during a page load. It is meant to be an approximate measure of the bandwidth usage of trackers. Expressed in kilobytes (KB).
    \textbf{Important note}: The distribution of this variable can have a fat tail due to audio or visual content sometimes served by third-party tracker requests. & Positive floating point\\
    \hline
    requests & Average number of third-party requests made to this company’s tracker per page load. & Positive floating point\\
    \hline
    requests\_tracking & Average number of third-party requests that contains potentially identifying information (cookie or query string) made to this company’s tracker per page load. & Positive floating point\\
  \bottomrule
\end{tabular}
\end{table*}

Karaj et al. describes how they operationalized the capture and aggregation of trackers’ data \cite{karaj_whotracks_2019}. There are three important concepts necessary to understand what this data represents.

First, a \textbf{page load} is defined by:

\begin{itemize}
\item Creation in the main web request of a tab when entering a URL in a web browser’s URL bar.
\item Ending when the tab is closed or another main web request is observed for the same tab, usually occurring when a link is clicked and the browser loads a new page.
\end{itemize}

Second, \textbf{third-party requests} are the building blocks of a tracker \cite{yu_tracking_2016}. During a page load, subsequent requests to a URL on a different domain than the current loaded page are counted as a third-party request.

Third, \textbf{aggregation} is possible with Ghostery’s trackerdb\footnote{\url{https://github.com/ghostery/trackerdb}}, a manually curated database mapping domain names to the companies they operate under. Using trackerdb’s information, an aggregation of data by companies is possible by linking third-party requests to specific domains and their parent company.

\subsection{SEC Filings}
Publicly traded companies in the United States have an obligation to comply with the Securities and Exchange Commission (SEC). This requires them to submit standardized forms and reports which are then published through the SEC’s Electronic Data Gathering, Analysis, and Retrieval (EDGAR) system. As this study focuses on GAFAM, all publicly traded companies' financial data is publicly available. Quarterly and annual earnings reports, published as part of the 10-Q and 10-K forms, provide us with insight into the companies' finances and their evolutions. 

Fundamental financial metrics were systematically extracted from these quarterly reports. Our analysis is based on four core financial metrics, from which two additional measures were derived, as shown in Table \ref{tab:finvars}.

\begin{table}
  \caption{The financial metrics collected and derived from 10-Q and 10-K SEC filings}
  \label{tab:finvars}
  \begin{tabular}{|p{0.25\linewidth} | p{0.65\linewidth}|}
    \toprule
    Financial Metrics&Description\\
    \midrule
    \hline
    Gross Revenue & The total amount of money earned in a quarter (in millions).\\
    \hline
    Advertising Services Revenue & The amount of money earned by the company through its advertising-related services (in millions).\\
    \hline
    Total Expenses & The total amount of money spent during a particular quarter (in millions).\\
    \hline
    Sales and Marketing Expenses & The amount of money spent in sales and marketing (in millions).\\
    \hline
    Share of Advertising Revenue & The proportion of the total revenue coming from advertising services (in percent).\\
    \hline
    Share of Marketing Expenses & The proportion of the total expenses dedicated to sales and marketing (in percent).\\
  \bottomrule
\end{tabular}
\end{table}

\subsection{Visual Analysis}
To grasp the GAFAM's role within the space of web tracking and their role in Surveillance Capitalism, visual analysis techniques were used to highlight patterns and trends. The analysis examined the progression of the tracking and financial data from May 2017 to March 2025.

This study used Python and Jupyter notebooks for data aggregation and visualization processing. Two additional Python libraries for data manipulation and representation were used: Pandas and Matplotlib. All code used in this study will be released in the public domain to facilitate replication and extension of this work (\textit{see} Appendix A).

\section{Findings}
Our analysis reveals significant patterns regarding GAFAM tracking practices. To contextualize the overall web tracking landscape, several non-GAFAM companies demonstrate substantial reach or site reach when averaged across the full temporal range (see Table \ref{tab:topreach}). Two of them, Twitter/X and Kaspersky Lab, are not publicly traded but have an estimated quarterly gross revenue of approximately 0.25B USD. Cloudflare and ComScore are publicly traded, and their latest quarterly gross revenues (Q4 2024) are 0.46B USD and 0.90B USD. These figures are eclipsed by the earnings of the GAFAM, with Facebook having the lowest revenue among GAFAM at 48.38B USD for Q4 2024. 46.78B USD of the 48.38B USD (96.69\%) came from advertising alone. The five companies examined in this study were chosen because these companies' yearly revenues rival the GDP of medium-sized nations, such as Finland, Greece, and Portugal\footnote{\url{https://data.worldbank.org/indicator/NY.GDP.MKTP.CD?locations=GR-FI-PT}} as shown in Table \ref{tab:gdp}.

\begin{figure*}[h]
  \centering
  \includegraphics[width=\linewidth]{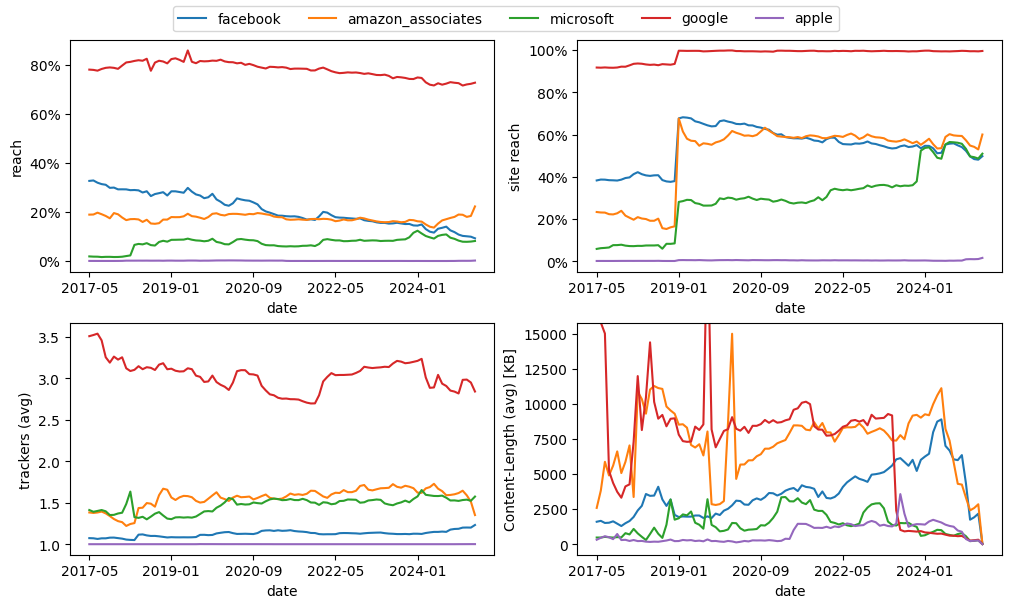}
  \caption{Historical evolution of GAFAM's web tracking}
  \Description{Four graph showcasing Google's dominance in the web tracking space.}
  \label{fig:gafamtrack}
\end{figure*}

\begin{table}[h]
  \caption{Top 7 companies based on average reach and site reach over the full time period}
  \label{tab:topreach}
  \begin{tabular}{|c|rl|rl|}
    \toprule
    Rank&Company&Reach&Company&Site Reach\\
    \midrule
    1&\textbf{Google}&\textbf{77.86\%}&\textbf{Google}&\textbf{98.02\%}\\
    2&\textbf{Facebook}&\textbf{21.01\%}&Kaspersky Lab&61.81\%\\
    3&\textbf{Amazon}&\textbf{17.46\%}&\textbf{Facebook}&\textbf{54.37\%}\\
    4&Cloudflare&7.72\%&\textbf{Amazon}&\textbf{50.33\%}\\
    5&\textbf{Microsoft}&\textbf{7.41\%}&Cloudflare&32.37\%\\
    6&Twitter/X&7.00\%&\textbf{Microsoft}&\textbf{29.69\%}\\
    7&ComScore&6.70\%&Twitter/X&29.49\%\\
  \bottomrule
\end{tabular}
\end{table}
\newpage
Two significant observations emerge from Table \ref{tab:topreach}: Google dominates the rankings both in terms of tracker reach and site reach, but Apple is notably absent. Figure \ref{fig:gafamtrack} demonstrates that Apple is the GAFAM company that tracks the least on the web. 

\begin{table}[h]
  \caption{2023 Annual Revenue/GDP (in billions)}
  \label{tab:gdp}
  \begin{tabular}{|cc|}
    \toprule
    Company/Country&Revenue/GDP\\
    \midrule
    \textbf{Amazon}&\textbf{574.78 USD}\\
    \textbf{Apple}&\textbf{383.28 USD}\\
    \textbf{Google}&\textbf{307.39 USD}\\
    Finland&295.53 USD\\
    Portugal&289.11 USD\\
    Greece&243.50 USD\\
    \textbf{Microsoft}&\textbf{211.91 USD}\\
    \textbf{Facebook}&\textbf{134.90 USD}\\
  \bottomrule
\end{tabular}
\end{table}

It is important to clarify that, in Figure \ref{fig:gafamtrack}, the spike in site reach shown in February 2019 is due to a change in how site reach is measured.\footnote{Refer to Table \ref{tab:vars} for an explanation about the change in methodology} As expected from Table \ref{tab:topreach}, Google dominates the realm of web tracking, with trackers present on almost 100\% of the top 10,000 websites visited by the users of the Ghostery extension. 

The reach, site reach, and average number of trackers data suggest a three-tier stratification among GAFAM in the web tracking space:

\begin{enumerate}
    \item Google is leading by a large margin.
    \item Facebook, Microsoft, and Amazon seem to be competing for the remaining surveillance opportunities.
    \item Apple appears strategically absent, possibly recognizing asymmetric competitive conditions and focusing their surveillance activities in alternative spaces.
\end{enumerate}
    
Another relevant tracker metric is the Content-Length average of third-party tracking requests. As shown in Figure \ref{fig:gafamtrack}, this metric presents findings that do not exhibit particular trends. These patterns suggest that this data should be considered cautiously. As specified in the documentation of the dataset’s variables, the Content-Length serves as an approximation. Empirical conclusions cannot be drawn on the accurate amount of bandwidth used by GAFAM’s user tracking. However, presenting this data is important to showcase that trackers incur a cost borne by users \cite{hanson_tracker_2018}. 
    
Three surprising patterns emerge from the data:
    
First, Google's and Amazon’s trackers seem to consume as much as 10MB of data on average when loading a page that contains their trackers. While caching would mitigate this to some extent, 10MB is very large for web content. The distribution of this data is supposed to have a fat-tail towards high Content-Length because some of the tracker domains also serve audio and video content \cite{karaj_whotracks_2019}. Averaging data with such a distribution will inflate the average, thus explaining this surprisingly large Content-Length.
    
Second, there is a sharp drop in Google trackers’ Content-Length in June 2023. Upon closer inspection of the dataset, it became apparent that it was primarily due to the “Youtube” tracker which corroborates the assumptions for the first pattern. This fall could be explained by an update to their tracking software that would stop reflecting the Content-Length of audio and visual content. Interestingly, in June 2023, Google's Privacy Sandbox initiative announced a change in their Topics API that is explicitly described as reducing data. Google present the Topics API as “designed to enable websites to serve relevant ads in a privacy-preserving manner, without resorting to covert tracking techniques, like browser fingerprinting. Topics utilizes several techniques to preserve user privacy, \textbf{including reducing data} [emphasis added], [...].”\footnote{\url{https://privacysandbox.google.com/blog/topics-enhancements\#update_june_15_2023} - \href{https://web.archive.org/web/20250410200834/https://privacysandbox.google.com/blog/topics-enhancements\#update_june_15_2023}{Internet Archive link}}
    
\begin{figure*}[h]
  \centering
  \includegraphics[width=\linewidth]{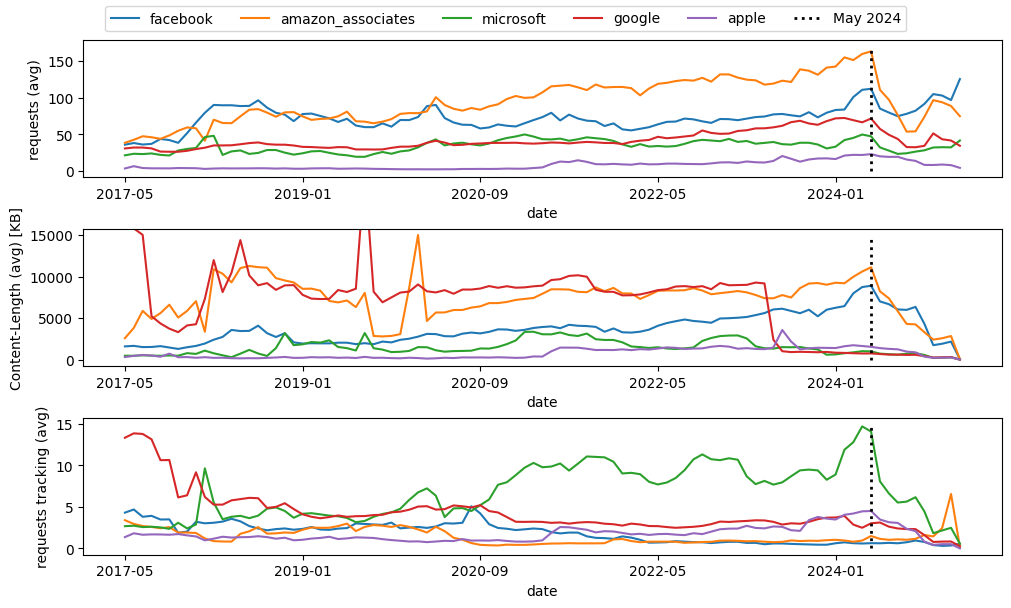}
  \caption{Historical evolution of third-party trackers requests}
  \Description{The amount of data and requests tracking users trending toward zero starting May 2024.}
  \label{fig:requests}
\end{figure*}

Third, starting in May 2024, all the trackers’ Content-Length trends toward zero. It seems improbable that all the GAFAM companies suddenly decided to reduce the bandwidth usage of their trackers at the same time. Rather, this pattern suggests a more fundamental shift in the way web tracking operates, showcasing the web tracking space as ever evolving, which further adds to its opacity. This shift is made obvious by Figure \ref{fig:requests} which highlights a decrease in the average number of requests and an even clearer trend in requests detected as containing identifying information. May 2024, denoted by the black dotted line in Figure \ref{fig:requests}, corresponds to Google's Privacy Sandbox announcement at Google I/O 2024 about third-party cookie deprecation in Google Chrome\footnote{\url{https://privacysandbox.google.com/blog/google-io-2024} - \href{https://web.archive.org/web/20250424012221/https://privacysandbox.google.com/blog/google-io-2024}{Internet Archive link}} further corroborating our assumption of an evolving web tracking landscape.

There is a high likelihood that the second and third patterns are related as they correspond to two announcements concerning the same products made by the same entity within Google (Privacy Sandbox)\footnote{\url{https://privacysandbox.com} - \href{https://web.archive.org/web/20250601173608/https://privacysandbox.com/}{Internet Archive link}}. Those two successive patterns bring to light another clue hinting at Google being the unequivocal leader of web tracking: Google first announced a change in the Chrome web browser API that is immediately implemented on their services (notably Youtube). That change is then advertised and promoted through an announcement at Google I/O a year later, which forced all the other GAFAM to follow suit and gradually adopt that change. One could even argue that Google is the architect of the web tracking landscape.

\begin{figure*}[h]
  \centering
  \includegraphics[width=\linewidth]{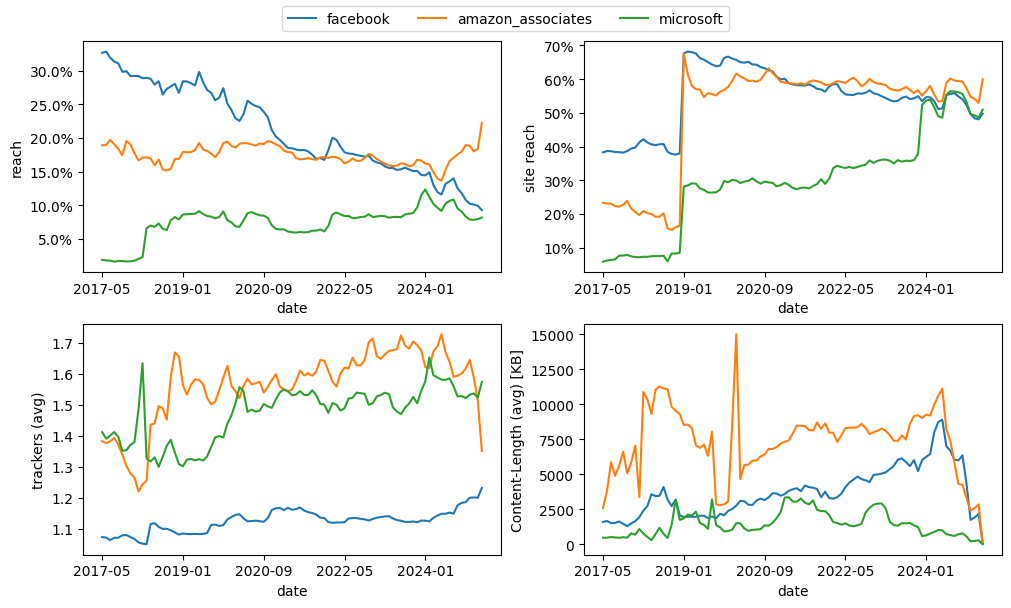}
  \caption{Historical evolution of "FAM's" web tracking}
  \Description{Four graph showcasing Facebook, Amazon and Microsoft's historical tracking data.}
  \label{fig:fam}
\end{figure*}

Figure \ref{fig:fam} focuses on the second-tier companies of the web tracking space, namely Facebook, Microsoft, and Amazon. Here, Facebook demonstrates a steady declining reach and site reach while simultaneously increasing the number of trackers and Content-Length size. This potentially indicates intensification strategies to maximize data extraction despite a declining presence. Microsoft exhibits slow but gradual increases in reach and site reach, with a particularly sharp increase in site reach in December 2023. This likely reflects Microsoft's strategic positioning in the artificial intelligence space and their partnership with OpenAI, requiring expanded data acquisition for model training.

\begin{figure*}[h]
  \centering
  \includegraphics[width=\linewidth]{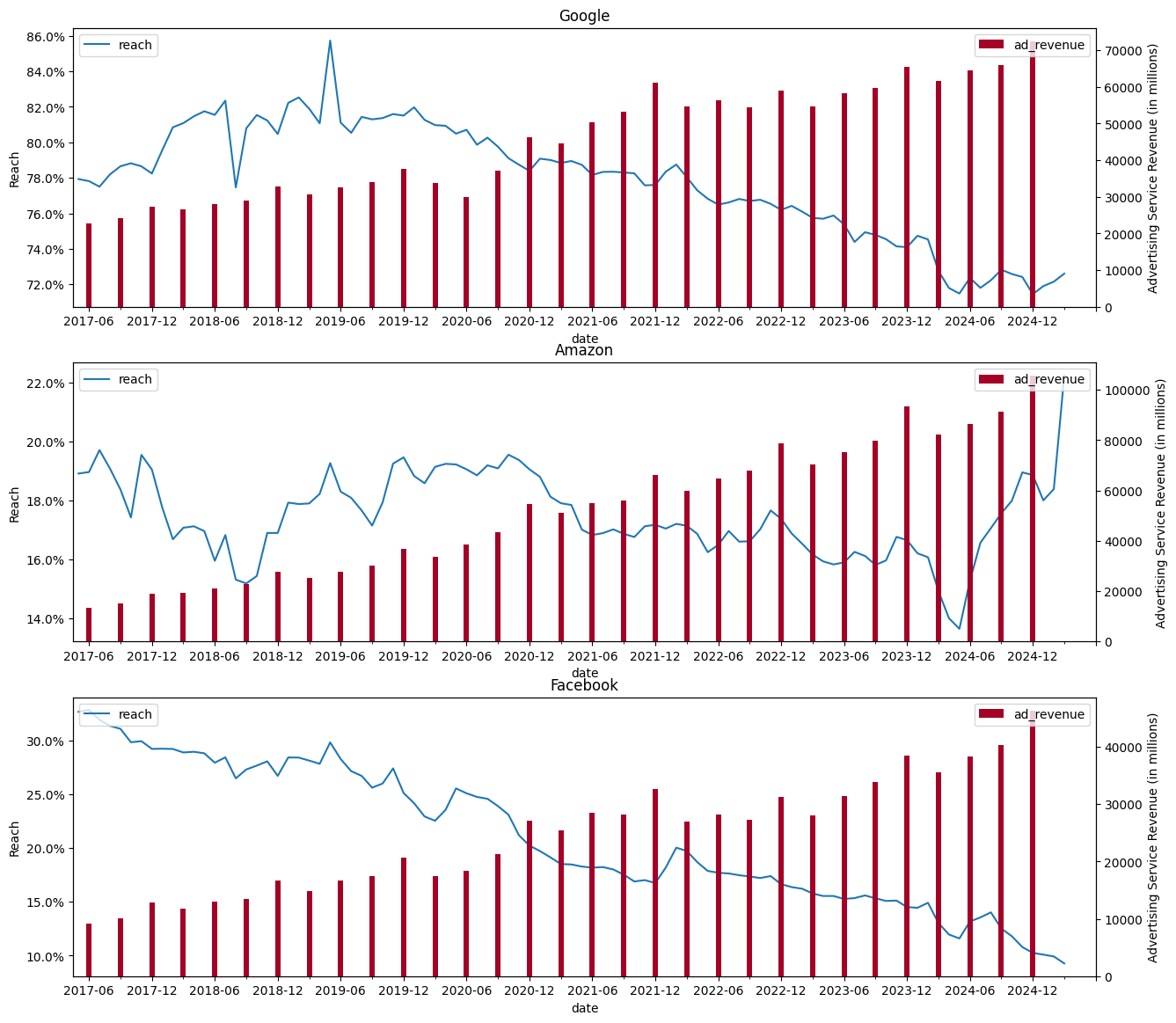}
  \caption[hello]{Google, Amazon and Facebook's quarterly advertising revenue overlayed on top of historical reach\textsuperscript{*}}
  \small\textsuperscript{*} Note: The y-axes are all on different scales as we mostly interested by the trends
  %\Description{Four graph showcasing Facebook, Amazon and Microsoft's historical tracking data.}
  \label{fig:googleadrevenue}
\end{figure*}

Recognizing that Google is the dominant player in surveillance, Figure \ref{fig:googleadrevenue} presents the company's advertising revenue growth overlaid on their tracker reach. A plausible interpretation of the inverse correlation suggests an increased efficiency in advertising revenue extraction despite the reduced user reach. The rise in reach until 2020, where the trend reverts, may indicate a “critical mass” of data has been acquired which enabled optimization of advertising revenue generation without a corresponding increase in tracking reach. Furthermore, looking at the three companies\footnote{Both Microsoft and Apple were omitted because extracting advertisement revenue from their SEC filings presented significant challenges discussed in the Limitations section.} showcased in Figure \ref{fig:googleadrevenue}, the graphs do not appear to show any correlation between tracking reach and ad revenue. This suggests that changes in tracking reach, as measured in the WhoTracks.Me dataset, do not have a direct impact on the companies' advertisement revenue. It is also worth noting that the impact of the Covid-19 pandemic can only be seen in the financial data of Google.

\section{Discussion}
Web tracking represents an integral part of Surveillance Capitalism's new economic order. As Zuboff argues, 'Big Other' --- her term for the expression of power produced by the uncontested global architecture of computer mediation essential to Surveillance Capitalism --- is constituted by mechanisms of extraction, commodification, and control. Web tracking is a materialization of this first step of extraction. With this paper's findings, it becomes clear how “High tech firms, led by Google, perceived new profit opportunities in these facts. Google understood that were it to capture more of these data, store them, and analyze them, they could substantially affect the value of advertising. As Google’s capabilities in this arena developed and attracted historic levels of profit, it produced successively ambitious practices [...].” \cite{zuboff_big_2015}. The difficulty in accurately capturing tracking activities is the foundation upon which the power imbalance of Surveillance Capitalism is built. In essence, web tracking functions as a one-way mirror where GAFAM learn about users and their habits, while providing little to no transparency regarding what kind of information they extract, where they store it, and how they leverage it for profit.

\subsection{Apple \& Google}

In Figure 1, Apple displays the least amount of web tracking activity. This seems consistent with Apple’s publicly held stance of defending the privacy rights of their customers. However, the recent class action lawsuit settlement regarding Apple’s Siri eavesdropping on their user raises questions about that stance.\footnote{\url{https://www.cbc.ca/news/business/apple-siri-privacy-settlement-1.7422363} - \href{https://web.archive.org/web/20250105085912/https://www.cbc.ca/news/business/apple-siri-privacy-settlement-1.7422363}{Internet Archive link}} Even more so, in the age of Surveillance Capitalism, it seems implausible for a corporation of Apple's scale not to leverage its customers' metadata for profit generation. A plausible hypothesis, warranting further investigation, suggests that Apple conducts surveillance through alternative channels rather than the web. An example of such an alternative channel is the App Store. Each time an Apple device searches for, comments on, rates, or installs an app, it sends requests to Apple's servers, leaving behind a significant amount of valuable digital traces.

This may explain Apple’s resistance to the integration of progressive web-apps (PWAs), a feature of web browsers introduced by Google\footnote{PWAs are a feature of web browsers that allow users to install a web-app/website very similarly to a native app. This has the side effect of bypassing the various app-stores altogether, effectively promoting even further the web from a space to share document, to an application platform.}, in their mobile operating systems. It is a strategy potentially designed to maintain users within the confines of the Apple’s closed ecosystem rather than letting them venture into Google’s surveillance territory. This understanding could also explain the controversial stance of Apple advocating for their customer’s privacy as a right. Apple’s threat model is obscure, they never disclosed to their users against who they would protect their privacy and against who they would not.

Though Apple is the least active in the web tracking space, Google is the key player. As Zuboff explains, Surveillance Capitalism originated at Google, making their dominant position unsurprising. An ex-Googler that worked on the Google Chrome team claimed that "the web is what browser vendors ship, you know, that's just the reality".\footnote{\url{https://www.localfirst.fm/2} - \href{https://web.archive.org/web/20250123203700/https://www.localfirst.fm/2}{Internet Archive link}} This type of testimony, along with documented empirical evidence of significant web tracking by Google and its latest attempt at overtaking the web tracking landscape with its Privacy Sandbox initiative\footnote{\url{https://www.theverge.com/2021/3/16/22333848/google-antitrust-lawsuit-texas-complaint-chrome-privacy} - \href{https://web.archive.org/web/20250607151912/https://www.theverge.com/2021/3/16/22333848/google-antitrust-lawsuit-texas-complaint-chrome-privacy}{Internet Archive link}}, reveals a fundamental corporate strategy: transforming the open web into a product to generate profit from.

\subsection{Cost of Surveillance Capitalism}
To quote Zuboff \cite{zuboff_big_2015}, for Google (and other GAFAM companies), “What matters is quantity not quality... Google is ‘formally indifferent’ to what its users say or do, as long as they say it and do it in ways that Google can capture and convert into data.” Such an extractive approach is particularly concerning since Surveillance Capitalism's carries both social and environmental costs. GAFAM’s ‘formal indifference’ enables their obsession to accumulate data for commodification and profit without concern for the damages to humans, society, or the environment.

Web trackers enable companies to measure and optimize various metrics (e.g., views, clicks, user engagement, conversion rate, impressions, etc.). This optimization, if left unchecked, can have dramatic social consequences. ‘Formally indifferent’ companies can and will trigger behaviors to generate data for capture and commodification. An example is the dissemination of increasingly politically divisive and controversial content online \cite{mcloughlin_misinformation_2024, ribeiro_microtargeting_2019}. The economic value of behavioral data derives from the ability to influence consumer behavior. With this power in the hands of ‘formally indifferent’ corporations, it has been used to increase consumerism without concern for the environmental burden it has placed on our already destabilized ecosystems and limited resources.

This investigation revealed web tracking’s non-zero bandwidth cost \cite{hanson_tracker_2018}. Given more precise and reliable measurements, this bandwidth could be linked to CO2 emissions \cite{ficher_assessing_2021, moulierac_what_2023}. Beyond transmitted tracking data, the surveillance apparatus enables targeted advertising and encourages sophisticated online advertising campaigns against web users. Far from innocuous, ads incur bandwidth and additional power consumption costs \cite{khan_impact_2024}.

\subsection{An Alternative to the Surveilled Web}
Though Surveillance Capitalism is omnipresent, various online communities\footnote{\url{https://smolweb.org} - \href{https://web.archive.org/web/20250419034754/https://smolweb.org/}{Internet Archive link}} and movements\footnote{\url{https://indieweb.org} - \href{https://web.archive.org/web/20250426150405/https://indieweb.org/}{Internet Archive link}} have taken an active stance against Surveillance Capitalism. As much as those communities deserve attention, a more radical alternative needs to be highlighted, one that leaves behind most modern web technologies --- JavaScript, CSS, Cookies, and other elements essential to the digital surveillance economy --- while preserving the web's most essential function: sharing and browsing hyperlinked documents. There exists alternative, non-extractive pathways that can serve as blueprints to share and engage with content on the internet.

Gemini\footnote{\label{note:gemini}\url{https://geminiprotocol.net} - \href{https://web.archive.org/web/20250327231005/https://geminiprotocol.net/}{Internet Archive link}} is described as “a new internet technology supporting an electronic library of interconnected text documents.” The Gemini protocol is similar in function to HTTP but with a limited subset of functionalities centered around document sharing. Together with the Gemtext format --- a simple markup language for Gemini similar in function to HTML for the web --- they form the foundation of a small but active alternative network. Gemini's lightweight nature and limited features foster an ecosystem of free and open-source software promoting simplicity, transparency, sharing, and learning. Importantly, Gemini's radical departure from the web makes its content inaccessible from mainstream browsers unless translated and mirrored to web technologies (possible due to Gemtext’s simplicity). This pseudo-isolation combined with technological simplicity has enabled a subculture and community to part ways with the corporate web and its extractive and harmful surveillance practices. The narrow feature-set provided by Gemini limits corporate interests and creates a ``[...] lightweight online space where documents are just documents, in the interests of every reader's privacy, attention and bandwidth.''\footref{note:gemini}

\section{Limitations}
Several limitations are present in this paper. Most of the data used in this investigation was sourced from WhoTracks.Me. Ghostery is a privacy-enhancing extension that cannot freely collect information on its users. This feature prevents systemic control for sample bias. In fact, evidence of such biases can be seen in the popularity ranking of websites within the data. For March 2025, fiverr.com was ranked as the most popular website globally, placing google.com second. For December 2023, the fifth most popular website was loot.tv. These anomalies nonetheless do not invalidate the whole dataset as the rest of the website rankings are relatively consistent with rankings coming from established sources such as Semrush\footnote{\url{https://www.semrush.com/website/top/}} or Similarweb\footnote{\url{https://www.similarweb.com/top-websites/}}. There is also no reason to believe that these irregularities disrupt longitudinal trends.

Furthermore, as demonstrated in the findings, the WhoTracks.Me data cannot be relied upon to accurately measure the amount of data consumed by web tracking. The data instead suggests that web tracking relies on data (such as scripts, cookies, images, etc.) transmitted via HTTP requests and this data consumes bandwidth. The cost of web tracking is non-zero and warrants further study with more reliable metrics to establish a proper lower bound on the data consumption of web tracking.

A second limitation relates to the dataset’s exclusive account of page loads containing trackers. If a website does not track its users, it will not appear in the dataset. This forecloses the analysis to the possibility of discovering potential alternatives or “ways-out” of the Surveillance Capitalism apparatus. Though, looking at the big picture, all top 50 most frequented websites track their users. This suggests minimal impact on the findings of this investigation.

There are inherent limitations to using financial data from the quarterly and annual earnings reports. Unfortunately, access to comprehensive financial data typically requires the use of proprietary or subscription-based services. Performing manual collection of the financial data from 10-Q and 10-K filings limited our analysis to a subset of the financial metrics. 10-Q and 10-K filings do not have a standardized format across companies or even across years within the same company's filings, making specific data, such as ad revenue, challenging to extract. For instance, Microsoft changes the definition of its advertisement category. This is reflected in the name of the line item that has changed over the years from “Advertising” to “Search advertising” to “Search and news advertising”, permitting inconsistent reporting of ad revenue. In the case of Apple, the reporting of ad revenue is hidden within a broader “Services” category which reports Advertising, AppleCare, Cloud Services, Digital Content, and Payment Services revenues under a single number.

Lastly, Surveillance Capitalism is a large and systemic phenomenon characterized by complex system interactions. Although web tracking provides valuable insights into a concrete mechanism of surveillance and monetization, these findings are, however, limited only to one category of data extraction. Surveillance Capitalism encompasses many additional aspects: data storage and processing, prediction services derived from that data, and infrastructure designs enabling various extraction and manipulation practices, such as dark patterns and digital rights management. The combination of web tracking and SEC filing data can only provide some clues about a small part of this broader socioeconomic phenomenon. It does not capture how the tracking data is then processed and used, nor does it provide insights into potential harms, such as undermining democratic processes or diverting attention away from the environmental consequences of unbridled neoliberalism.

\section{Conclusion}
This empirical investigation provides valuable insights into how Surveillance Capitalism is operationalized by the world's most powerful technology companies. Our longitudinal analysis of tracking data from May 2017 to March 2025 reveals patterns that confirm and extend our theoretical understanding of the extent of surveillance on the web.

Google was identified as the dominant surveillance entity on the web, with tracking reach largely exceeding the other GAFAM companies. The analysis of our results surfaced a three-tiered stratification among the GAFAM companies, with Google at the top, Facebook, Amazon, and Microsoft competing in a second tier, and Apple strategically absent. This stratification hints at how surveillance practices reflect and reinforce capital market power dynamics.

The relationship between tracking reach and advertising revenue, particularly in Google's case, suggests that Surveillance Capitalism may be entering a new phase characterized by more efficient and aggressive data exploitation rather than merely increasing the scale of collection. This evolution has roots in reality, as Google’s trackers are already present on almost 100\% of the top websites. Further expansion is unfeasible forcing Google to innovate ways to extract behavioral predictive data out of a “data exhaust” that already reached its maximum.

This study confirms that web tracking data can serve as concrete empirical evidence of Surveillance Capitalism. By quantifying surveillance activities, through web tracking, and connecting them to financial metrics, we move beyond qualitative frameworks to measurable phenomena.

Several promising directions for future research emerge from this investigation. Expanding the historical scope beyond 2017 could provide an understanding of the continued evolution of Surveillance Capitalism. The approach suggested by Lerner et al. to leverage the Internet Archive's Wayback Machine to identify historical trackers offers a promising approach \cite{lerner_internet_2016}. Apple's apparent lack of presence in web tracking warrants closer examination of how Surveillance Capitalism operates within closed ecosystems. Recognizing how surveillance can manifest under seemingly privacy-respecting systems can reveal different forms of data extraction and behavioral surplus generation. Developing more sophisticated methods for extracting and analyzing financial data in relation to surveillance activities --- potentially through automated extraction and analysis of SEC filings --- can also surface new insights into the economic mechanisms of Surveillance Capitalism. In particular, ad revenue may be related to increasingly manipulative practices to engage users with politically divisive content. This phenomenon deserves greater attention, especially with political polarization increasing within Western democracies \cite{mccoy_polarization_2018}. Finally, comparative studies examining impacts of jurisdictional differences, particularly between the EU and US, on web tracking and financial metrics could illuminate how regulatory environments shape surveillance practices and enable or limit surveillance-enabled economic growth.

As web technologies continue to evolve with tracking practices, ongoing empirical monitoring of surveillance practices remains essential to understand both the current and future trajectory of Surveillance Capitalism. This study contributes to that effort by demonstrating the value of web tracking data as a window into the actual mechanisms through which our collective online experience is captured and commodified.
    
\begin{acks}
Christoph for inspiring this investigation and for his guidance.\\
Sayaka for her unwavering support and advice.
\end{acks}

\bibliographystyle{ACM-Reference-Format}
\bibliography{references}

\appendix

\section{Source Code}
\begin{table}[h]
  \caption{Software Tools}
  \label{tab:tools}
  \begin{tabular}{|c c c|}
    \toprule
    Name&Version&Website\\
    \midrule
    Python&3.12.4&\url{https://www.python.org}\\
    Jupyter Lab&4.3.4&\url{https://jupyter.org}\\
    Pandas&2.2.3&\url{https://pandas.pydata.org}\\
    Matplotlib&3.10.1&\url{https://matplotlib.org}\\
  \bottomrule
\end{tabular}
\end{table}

The code is published on the author's own website\footnote{\url{https://fsl.blazebone.com/empirical_inquiry_into_sc_limits_2025_source_code}} in the form of a Fossil repository in addition to an archive of the code in the Internet Archive\footnote{\url{https://archive.org/details/empirical_inquiry_into_sc_limits_2025_analysis}}. GitHub has been considered but will be avoided as it currently belongs to Microsoft, one of the key actors of Surveillance Capitalism. Table \ref{tab:tools} lists the software tools along with their version that were used for the empirical analysis part of this study.

\end{document}